# Observation of Yu-Shiba-Rusinov-like states at the edge of $CrBr_3/NbSe_2$ heterostructure

Yuanji Li[1,†], Ruotong Yin[1,†], Mingzhe Li[1], Jiashuo Gong[1], Ziyuan Chen[1], Jiakang Zhang[1], Ya-Jun Yan[1,2,*], Dong-Lai Feng[1,2,3,*]

[1] *School of Emerging Technology and Department of Physics, University of Science and Technology of China, Hefei, 230026, China*

[2] *Hefei National Laboratory, University of Science and Technology of China, Hefei, 230088, China*

[3] *National Synchrotron Radiation Laboratory, School of Nuclear Science and Technology, and New Cornerstone Science Laboratory, University of Science and Technology of China, Hefei, 230026, China*

**\*Corresponding authors:** <u>yanyj87@ustc.edu.cn; dlfeng@ustc.edu.cn</u>

**†These authors contributed equally to this work.**

The hybrid ferromagnet-superconductor heterostructures have attracted extensive attention as they potentially host topological superconductivity. Relevant experimental signatures have recently been reported in $CrBr_3/NbSe_2$ ferromagnet-superconductor heterostructure, but controversies remain. Here, we reinvestigate $CrBr_3/NbSe_2$ by an ultralow temperature scanning tunneling microscope with higher spatial and energy resolutions. We find that the single-layer $CrBr_3$ film is insulating and acts likely as a vacuum barrier, the measured superconducting gap and vortex state on it are nearly the same as those of $NbSe_2$ substrate. Meanwhile, in-gap features are observed at the edges of $CrBr_3$ island, which display either a zero-energy conductance peak or a pair of particle-hole symmetric bound states. They are discretely distributed at the edges of $CrBr_3$ film, and their appearance is found closely related to the atomic lattice reconstruction near the edges. By increasing tunneling transmissivity, the zero-energy conductance peak quickly splits, while the pair of nonzero in-gap bound states first approach each other, merge, and then split again. These behaviors are unexpected for Majorana edge modes, but in consistent with the conventional Yu-Shiba-Rusinov states. Our results provide critical information for further understanding the interfacial coupling in $CrBr_3/NbSe_2$ heterostructure.

**Introduction**

The interplay of magnetism and superconductivity induces a wealth of novel quantum states, such as high-temperature superconductivity[1-3], Fulde-Ferrel-Larkin-Ovchinnikov (FFLO) state[4-6] and possibly, topological superconducting states[7-9]. The basic form of interplay between magnetism and superconductivity is the presence of individual magnetic impurities in a superconductor. The exchange coupling between the local spin and Bogoliubov quasiparticles of the superconductor leads to low-energy in-gap bound states known as Yu-Shiba-Rusinov (YSR) states[10-15]. The energies of the YSR states correspond to the level spacing between the ground state and the lowest excitation state of local spin, conforming to the formula: $\varepsilon = \Delta \frac{1-a^2}{1+a^2}$, where $\Delta$ is the superconducting gap size, $a = JS_{\text{imp}}\pi\rho_s$, with $S_{\text{imp}}$ being the spin state of magnetic impurity, $\rho_s$ being the density of state (DOS) at Fermi level ($E_F$) in the normal state, and $J$ being the exchange coupling strength between impurity spin and superconductor[15-20]. If magnetic impurities are arranged in high-density chains, individual YSR states will overlap and hybridize, possibly forming extended Shiba bands[7,8,15,21,22]. Electrons in such Shiba bands may hybridize with the condensate of bulk superconductor by Andreev reflection, whose strength depends on the magnetic structure of the impurity chain. Assuming a helical or ferromagnetic arrangement of impurity spins, it is argued that an effective topological superconducting phase, akin to one-dimensional (1D) spinless $p$-wave superconductors, can be realized, with Majorana zero modes (MZMs) residing at the ends of the magnetic chain[7,8,21,22]. Experimentally, topological Shiba bands were reported to be realized in Mn chains grown on Nb superconductor[23], evidence of MZMs were observed at the terminals of Fe chains adsorbed on Pb or Nb superconductors[24-27], but not for that grown on $NbSe_2$ (ref. 28). MZM was reported to be absent in diluted Cr spin chains grown on Nb superconductor[29], while enhanced zero-energy DOSs assigned as MZMs were reported at the ends of artificially hybrid transition-metal atom chains on Re superconductor[30], whose occurrence was tunable by the length and termination of the chains[31]. Furthermore, when magnetic adatoms are arranged in a two-dimensional (2D) lattice on the surface of a superconductor with strong spin-orbit coupling, a 2D topological chiral $p$-wave superconductor with 1D dispersive Majorana edge modes (MEMs) will be realized[32,33]. In experiments, signatures of chiral MEMs have been observed in Pb/Co/Si(111) (ref. 34), Fe/Re(0001)-O(2×1) (ref. 35), Cr/Nb(110) (ref. 36), Mn/Nb(110) (refs. 37, 38) heterostructures, but they are absent in $MnSe/NbSe_2$ (ref. 39) and $MnTe/NbSe_2$ (ref. 40). The main reason for these differences among different systems should be the finite phase space for the existence of topological superconductivity, which requires a delicate interplay among the magnetic, superconducting, and spin-orbital coupling of the heterostructures.

Recently, signatures of 1D chiral MEMs were reported in the heterostructure fabricated by ferromagnetic $CrBr_3$ monolayer film grown on $NbSe_2$, manifested as a zero-energy conductance peak (ZECP) distributed discontinuously along the edges of $CrBr_3$ island[41]. Being topologically protected, the MEMs are generally considered to be insensitive to lattice defects or disorders, thus they should be spatially continuous along the edges; moreover, as the 1D chiral MEMs are approximately linearly dispersive around zero energy, they should in principle fill up the superconducting gap and produce finite DOS at zero energy as the cases in refs. 32, 33, 35, but are not necessarily peaked at zero energy unless additional parameters are induced, such as a strong and spatially varying Zeeman field in Pb/Co/Si(111) (ref. 34). Therefore, the reported discrete ZECPs in $CrBr_3/NbSe_2$ heterostructure are beyond the conventional understanding of MEMs. To elucidate these anomalous phenomena, we construct $CrBr_3/NbSe_2$ heterostructures, and study them using scanning tunneling microscopy/spectroscopy (STM/STS). Benefiting to the much lower lattice temperature (40 mK) and higher energy resolution (~ 50 μeV that corresponds to an effective electron temperature $T_{eff}$ ~ 170 mK) of our STM system, as well as the higher spatial resolution of data acquisition, more valuable information is revealed. We find that the $CrBr_3$ film is insulating and acts as a vacuum barrier, the superconducting gap and vortex states measured on it are nearly identical to those of $NbSe_2$ substrate. Two types of edge states, one with a ZECP and the other with a pair of particle-hole symmetric in-gap bound states, are discretely distributed at the edges of $CrBr_3$ film. Their appearance is found to be closely related to the specific lattice reconstructions of the step edges. Moreover, tunneling transmissivity-dependent measurements find obvious evolution of these edge states as their exchange coupling strength $J$ varies, resembling the properties of conventional YSR states. Our results provide conclusive experimental evidence for the topologically trivial origin of edge states in $CrBr_3/NbSe_2$ heterostructure, and the detailed structural and electronic information help further understand the interfacial coupling effect.

**Results**

**Structural and electronic properties of $CrBr_3/NbSe_2$ heterostructure.** Figure 1a shows the typical topographic image of $CrBr_3/NbSe_2$ heterostructure, the $CrBr_3$ island is typically triangle-shaped with tens to hundreds of nanometers in size. The height of the CrBr$_3$ island is approximately 0.6 nm (Fig. 1b), close to the *c*-axis lattice constant of bulk CrBr$_3$ (ref. 42), confirming the monolayer-thick nature. Due to the in-plane lattice mismatch between NbSe$_2$ ($a = b = 3.44$ Å) and CrBr$_3$ ($a = b = 6.26$ Å), well ordered Moiré superstructure is formed, as shown in Fig. 1c. The period of the Moiré superstructure is ~ 6 nm, which may vary slightly due to a small change of the twisted angle between NbSe$_2$ and CrBr$_3$, as discussed in Supplementary Fig. 1 of Supplementary Information (SI). The fast Fourier transforms (FFT) of Fig. 1c is shown in its inset, exhibiting the Bragg spots of both the CrBr$_3$ lattice ($\mathbf{q}_a$) and Moiré superlattice ($\mathbf{q}_M$), with their orientations off by ~ 30°. Figure 1d shows the atomically resolved topographic image of a monolayer CrBr$_3$ film, revealing evenly spaced clover-shaped atomic patterns. These features are consistent with previous reports[41,43,44], which are contributed by the three top Br atoms (white balls in the overlaid CrBr$_3$ atomic structure in Fig. 1d).

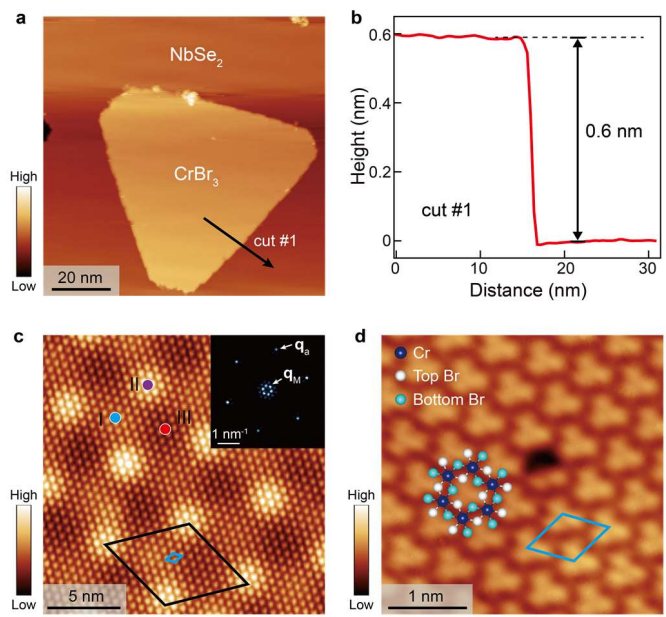

**Fig. 1 | Structural properties of CrBr$_3$/NbSe$_2$ heterostructure. a** Typical topographic image of monolayer CrBr$_3$ island grown on NbSe$_2$ substrate. **b** Line profile taken along cut #1 in (**a**). **c** Typical topographic image of CrBr$_3$ island. Hexagonal Moiré superlattice is clear, with its unit cell indicated by the black rhombus. Inset is its FFT image, with the Bragg spots of underlying CrBr$_3$ lattice and Moiré superlattice labeled as $\mathbf{q}_a$ and $\mathbf{q}_M$, respectively. **d** Atomically resolved topographic image of CrBr$_3$ film with an overlaid CrBr$_3$ atomic structure. The unit cell is indicated by the blue rhombus. Cr, top Br, and bottom Br atoms are represented by the blue, white, and cyan balls, respectively. Measurement conditions: (**a**) $V_b = 2$ V, $I_t = 20$ pA; (**c**) $V_b = 1.8$ V, $I_t = 30$ pA; (**d**) $V_b = 0.1$ V, $I_t = 150$ pA.

Then we study the electronic properties of CrBr$_3$ film, its typical d$I$/d$V$ spectra are shown in Figs. 2a and 2b, with those on bare NbSe$_2$ listed as well for comparison. When the stabilization bias voltage was set as $V_b = 1.7$ V, the spectra on CrBr$_3$/NbSe$_2$ and bare NbSe$_2$ differ significantly, the former behaves as an insulator with a large band gap between -2 eV and 1 eV (red curve in Fig. 2a), while the latter shows good metallic behavior (black curve in Fig. 2a). However, when $V_b = 0.9$ V that locates inside the insulating gap of CrBr$_3$, two nearly identical spectra are observed on CrBr$_3$/NbSe$_2$ and bare NbSe$_2$ as shown in Fig. 2b. This is because the CrBr$_3$ film has no local DOS in the measured energy range of Fig. 2b and behaves as a tunneling barrier, through which electrons tunnel into the underlying NbSe$_2$ layer. These results are consistent with previous reports[43-45], and similar phenomena have been widely observed by STM studies on insulator/superconductor or insulator/metal heterostructures, for examples, in MnTe/NbSe$_2$ (ref. 40) and CrI$_2$/NbSe$_2$ (ref. 46). Furthermore, when scrutinizing the spectral features around E$_F$, we find the spectrum on CrBr$_3$/NbSe$_2$ is shifted slightly towards higher energies by ~ 30 meV when compared to that of bare NbSe$_2$, as marked out in the inset of Fig. 2b, which implies a weak charge transfer at the CrBr$_3$/NbSe$_2$ interface. A similar band shift has been reported in previous studies on CrBr$_3$/NbSe$_2$ heterostructure[41].

Figures 2c,d and Supplementary Fig. 2 of SI show the spatial DOS modulation of CrBr$_3$ film near the NbSe$_2$ superconducting gap energies. The DOS is uniform at E = 0 meV (Fig. 2c), and is weakly modulated by the Moiré pattern at E = -0.72 meV (Fig. 2d). Figure 2e displays the superconducting gap spectra collected both on bare NbSe$_2$ and at different locations of the Moiré superlattice on CrBr$_3$ film, exhibiting essentially the same fully developed superconducting gap --- the coherence peaks locate at approximately ±1.3 meV and the gap bottom with nearly zero tunneling conduction is ~ 1 meV wide. Figures 2f,g are the enlarged views of the shaded regions shown in Fig. 2e. It is obvious that the gap bottoms of these spectra almost exactly coincide (Fig. 2g), and the deviation of gap size judging from the gap edge and coherence peak positions is less than 0.05 meV (Fig. 2f). Previous studies reported the existence of Moiré pattern modulated Shiba band in CrBr$_3$ film, manifested as in-gap DOS lift near ±0.3 meV (refs. 41, 47), which is not observed in our study. We just see the suppressed superconducting gap on a few defects in CrBr$_3$ film, as discussed in Supplementary Fig. 3 of SI.

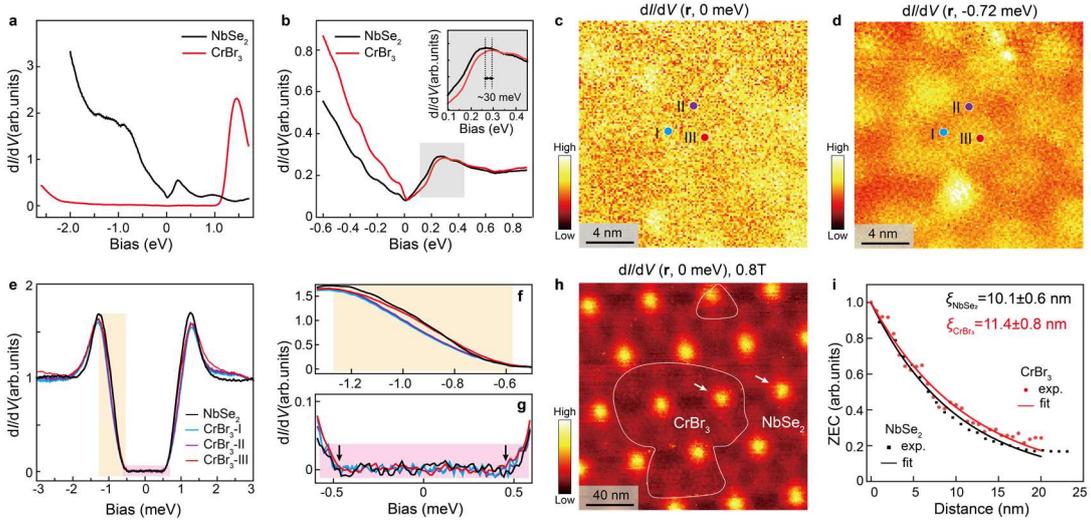

**Fig. 2 | Electronic properties of CrBr$_3$/NbSe$_2$ heterostructure. a,b** Typical tunneling spectra of the CrBr$_3$ film on NbSe$_2$ substrate and the bare NbSe$_2$ substrate. Inset of (**b**) shows the enlarged view of the shaded region. **c,d** d$I$/d$V$ maps on CrBr$_3$ film at E = 0 meV and -0.72 meV, respectively. The colored circles labeled as I, II, and III mark three different sites of Moiré superlattice, as also indicated in Fig. 1c. **e** Typical superconducting gap spectra taken on NbSe$_2$ and different locations of CrBr$_3$ film. Enlarged views of the shaded regions are displayed in (**f** and **g**). **h** Vortex map on CrBr$_3$/NbSe$_2$ heterostructure at E = 0 meV under **B**$_Z$ = 0.8 T. The regions where monolayer CrBr$_3$ film exists are enclosed by the white curves. **i** Exponential fits to the line profiles of free vortices located on CrBr$_3$ film and bare NbSe$_2$, as indicated by the white arrows in (**h**). Measurement conditions: (**a**) $V_b$ = 1.7 V, $I_t$ = 100 pA, $\Delta V$ = 40 mV; (**b**) $V_b$ = 0.9 V, $I_t$ = 200 pA, $\Delta V$ = 20 mV; (**c,d**) $V_b$ = 3 mV, $I_t$ = 200 pA, $\Delta V$ = 0.1 mV; (**e**) $V_b$ = 3 mV, $I_t$ = 200 pA, $\Delta V$ = 0.05 mV; (**h**) $V_b$ = 3 mV, $I_t$ = 100 pA, $\Delta V$ = 0.15 mV.

The superconductivity of CrBr$_3$/NbSe$_2$ heterostructure is further investigated by imaging magnetic vortices. Supplementary Fig. 4a of SI shows the topographic image of a selected sample region consisting of both CrBr$_3$ and NbSe$_2$, the corresponding zero-energy conductance (ZEC) maps under out-of-plane magnetic fields ($B_Z$) of 0.2 T and 0.8 T are displayed in Supplementary Fig. 4b of SI and Fig. 2h, respectively. Firstly, we find that the vortices form a perfect triangular Abrikosov lattice in the entire sample region, which suggests that the $CrBr_3$ islands and their edges have no significant pinning effect on vortex distribution. Secondly, the vortices on both $CrBr_3$ film and $NbSe_2$ substrate are essentially the same --- they are sixfold-symmetric star-shaped with high DOS tails and their vortex states both show the same "X"-shaped dispersion (Supplementary Fig. 4b-h of SI). These features are consistent with previous observations in bulk $NbSe_2$ (refs. 48-50). Thirdly, the estimated Ginzburg-Landau coherence length ($\xi$), extracted by an exponential fit to the line profile of each vortex under $B_Z$ = 0.8 T (Fig. 2i), is 10.1 ± 0.6 nm for bare $NbSe_2$ and 11.4 ± 0.8 nm for $CrBr_3/NbSe_2$, close to each other. These characteristics of electronic states, superconducting gap spectrum and magnetic vortex all suggest that the $CrBr_3$ film behaves likely as a vacuum barrier layer here, and it has little influence on the electronic properties of $NbSe_2$ underneath.

**Edge states of $CrBr_3$ film.** We then study the edge states at the boundary of $CrBr_3/NbSe_2$ heterostructure. Figures 3a,b show the typical topographic images of the edges of two $CrBr_3$ islands, which are imperfect and appear to be divided into relatively smooth segments of a few nanometers in length. The definition of smooth segments is marked out in Fig. 3a, and the region between them is called the junction area. The corresponding d$I$/d$V$ maps taken at E = 0 meV and ~ -0.4 meV are shown in Figs. 3c-f. In-gap states are observed at both edges, but their weights are mainly concentrated at the smooth segments of the edges, forming a spatially discontinuous distribution. Figure 3g shows the atomically resolved topographic image of a smooth segment of $CrBr_3$ edge, with an overlaid $CrBr_3$ atomic structure. Obviously, lattice reconstruction occurs at the outermost edge, resulting in a larger period of ~ 1.1 nm, which is approximately three times of the nearest neighbored Cr-Cr or Br-Br atomic spacing, as illustrated by the magenta curve in Fig. 3i. Moreover, we find that the distribution of edge states within a smooth segment is also discontinuous, exhibiting discrete bright spots modulated by the outermost lattice reconstruction, as shown in Figs. 3h,i. Please see Supplementary Fig. 5 of SI to find more datasets for the edge states. These results strongly indicate that the appearance of edge states in $CrBr_3$ film is closely related to the lattice reconstruction of the edges.

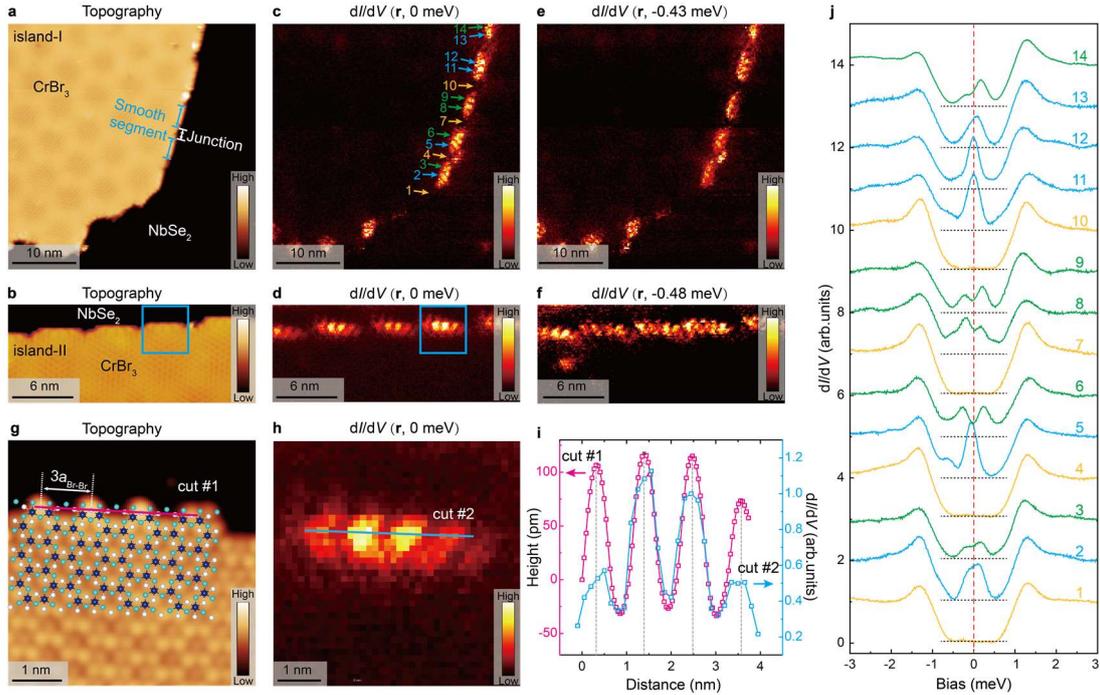

**Fig. 3 | Edge states of CrBr$_3$ film. a,b** Topographic images of CrBr$_3$ island-I and II. The definition of smooth segment and junction area is marked out in (**a**). **c-f** Corresponding d$I$/d$V$ maps of CrBr$_3$ island-I and II. **g** Atomically resolved topographic image of the edge of CrBr$_3$ island-II, collected in a smooth segment as indicated by the blue box in (**b**). The atomic structure of CrBr$_3$ is superimposed, with Cr, top Br, and bottom Br atoms represented by the blue, white, and cyan balls, respectively. **h** Enlarged image of the area indicated by the blue box in (**d**). **i** Line profiles taken along cuts #1 and #2 in (**g** and **h**), respectively. **j** Three kinds of representative d$I$/d$V$ spectra collected at different locations of the edges of CrBr$_3$ islands, as indicated by the color-coded arrows in (**c**). The horizontal black dashed lines indicate the positions of zero conductance for each spectrum. Measurement conditions: (**a**) $V_b$ = 0.9 V, $I_t$ = 30 pA; (**b**) $V_b$ = 0.9 V, $I_t$ = 20 pA; (**c,e**) $V_b$ = 3 mV, $I_t$ = 100 pA, $\Delta V$ = 0.15 mV; (**d,f**) $V_b$ = 3 mV, $I_t$ = 100 pA, $\Delta V$ = 0.1 mV; (**g**) $V_b$ = 0.33 V, $I_t$ = 30 pA; (**j**) $V_b$ = 3 mV, $I_t$ = 200 pA, $\Delta V$ = 0.05 mV.

Figure 3j shows the representative d$I$/d$V$ spectra collected at different locations of CrBr$_3$ edges, as indicated by the color-coded arrows in Fig. 3c. These spectra could be roughly classified into three types.

(i) The first type is a fully developed superconducting gap (orange curves in Fig. 3j), which is the same as that observed in bare NbSe$_2$ and usually appears in the junction areas between different smooth segments.

(ii) The second type shows a ZECP (blue curves in Fig. 3j), similar to previous reports[41]. However, the peak width of the ZECPs varies significantly, from 0.2 meV to 0.5 meV, and the peak shape is not always strictly symmetric with respect to $E_F$.

(iii) The last type shows a pair of particle-hole symmetric in-gap conductance peaks (green curves in Fig. 3j), which has not been reported before.

The latter two types of spectra are both observed in the smooth segments, but their distribution is random. Besides, we notice that there are ~ 13% of the spectra collected at the step edges that cannot be classified into the above three types, as indicated by the magenta curves in Supplementary Fig. 6 of SI, which are characterized by a single asymmetric in-gap peak or multiple pairs of in-gap peaks. As illustrated in Supplementary Fig. 7 of SI, the spatial locations of these spectra are usually close to defects/impurities or structural disorders, which will not be discussed in this manuscript. In ref. 41, the discretely distributed ZECP was considered as Majorana edge modes, as a result of Shiba bands coupling to an s-wave superconductor. However, in our case, the ZECPs that existed only at some localized positions of the step edges seem unlikely to be topologically protected, and the appearance of a pair of particle-hole symmetric in-gap states makes YSR states a more reasonable interpretation.

**Tunneling transmissivity-dependent d$I$/d$V$ spectrum measurements.** In the painstaking search of topological superconductivity and Majorana modes, ZECPs in local tunneling spectroscopy measurements are often caused by artifacts instead of MZMs, such as YSR states induced by local defects or by a different electronic structure at the chain terminations[28,29,51]. A more compelling evidence for a MZM is that it possesses the quantized conductance value of $G_0 = 2e^2/h$ due to resonant Andreev reflection, which could be observed at a finite temperature when the tunneling coupling is sufficiently strong[52,53]. Experimentally, the quantized conductance plateau has been observed in the free vortex cores of (Li,Fe)OHFeSe (ref. 54) and Fe(Te,Se) (ref. 55). Moreover, when changing the tunneling coupling strength, the MZM does not split and is strictly kept at zero energy due to the topologically protected nature.

To check this, we perform tunneling transmissivity-dependent measurements for the latter two kinds of edge states, the ZECP and the particle-hole symmetric in-gap conductance peaks. The transmissivity of tunneling barrier, defined as $G_N = I_t/V_b$, can be changed by controlling the tip-sample distance. We gradually increase the tunneling current $I_t$ at a fixed $V_b$ to reduce the tip-sample distance, and to increase $G_N$ (please see Supplementary Fig. 8 of SI to find the relationship between $I_t$ and $G_N$). During this process, the electrostatic force from the approaching tip will affect the coupling between the edge adatoms and the superconductor[56,57], leading to variable $J$ values. We find that with increasing $G_N$, the ZECP edge state quickly splits into two symmetric side peaks at $G_N \sim 0.008 \pm 0.002\ G_0$, which move to higher energies with further increasing $G_N$ (Fig. 4a). This process is reversible, the split peaks gradually return to a single ZECP when lowing $G_N$ again (Fig. 4b). Figures 4c,d show the corresponding waterfall plots for a clearer view, and Supplementary Fig. 9 of SI presents more datasets for the ZECP evolution. Meanwhile, the edge state of a pair of particle-hole symmetric in-gap conductance peaks also shows a strong dependence on $G_N$, as shown in Figs. 4e,f. For a very low $G_N$ value, the two side peaks are located at ~ ±0.2 meV, and the peak intensity at positive energy is stronger than that at negative energy (the bottom spectrum in Fig. 4e). By increasing $G_N$, they first approach to each other, merge into a ZECP at $G_N \sim 0.003 \pm 0.001\ G_0$ and then split again (Figs. 4e,g). For the re-split spectra, the intensity of side peaks at positive and negative energies is reversed compared with the initial spectrum. This process is also reversible when lowing $G_N$, as shown in Fig. 4f. Moreover, we notice that the spectral profiles and the critical $G_N$ values when increasing and lowering $G_N$ are slightly different (Figs. 4a,b and Figs. 4e,f), which may be related to the slight change of measurement sites due to the inevitable thermal drift effect during long period measurements.

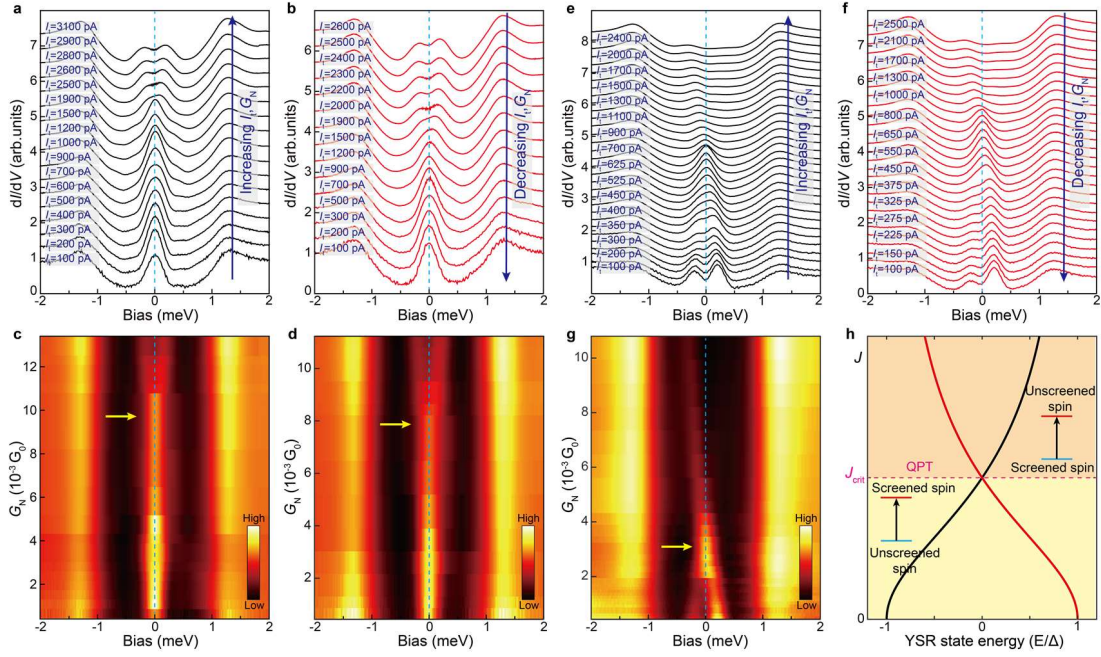

**Fig. 4 | Tunneling transmissivity dependence of two types of edge states. a,b** Evolution of the spectra with a ZECP as a function of $G_N = I_t/V_b$ with a fixed $V_b$ of 3 mV, obtained by increasing or decreasing $I_t$, respectively ($I_t$ = 100 - 3100 pA, $\Delta V$ = 0.05 mV). **c,d** Waterfall plots of (**a** and **b**). The yellow arrows indicate where the ZECP begins to split. **e,f** Evolution of the spectra with a pair of in-gap particle-hole symmetric conductance peaks as a function of $G_N$, obtained by increasing or decreasing $I_t$, respectively ($I_t$ = 100 - 2600 pA, $\Delta V$ = 0.05 mV). **g** Waterfall plot of (**e**). The yellow arrow indicates the location of the QPT point. **h** Sketched phase diagram of the evolution of YSR state versus $J$. The yellow region ($J < J_{crit}$) corresponds to the unscreened ground state of local spin, while the orange region ($J > J_{crit}$) corresponds to the Kondo-screened ground state. The magenta dashed line corresponds to the occurrence of QPT.

The spatial and energy distributions of these edge states and their tunneling transmissivity-dependent evolution cannot be explained by trivial edge states, nor do them conform to the expectation of MZMs, which shall be robust against perturbation. Instead, they are more likely the behavior of YSR states. According to the formula describing the energy positions of YSR states[15]: $\varepsilon = \Delta \frac{1-a^2}{1+a^2}$ with $a = JS_{imp}\pi\rho_s$, the evolution of YSR states versus $J$ is illustrated in Fig. 4h. For a small $J$, the system has an unscreened ground state, while for a large $J$ the system has a Kondo-screened ground state. At a critical $J_{crit}$, a quantum phase transition (QPT) occurs. In excitation spectra, with increasing $J$, a pair of particle-hole symmetric YSR states gradually shift to $E_F$, form a ZECP at $J_{crit}$, then split again. Across QPT, the relative spectral weight of the electron- and hole-like components of the YSR pair is reversed. These phenomena have been extensively studied in theory[14] and experiments[57-62]. In our study, comparing to Fig. 4h, the evolution of the nonzero in-gap conductance peaks strongly resembles the properties of topologically trivial YSR states that undergo a QPT, while the ZECP is also YSR states that located coincidentally at the QPT point. Moreover, the nonzero in-gap conductance peaks are very sensitive to the changes of $G_N$ (Fig. 4g), while the ZECP keeps unchanged for a larger range of $G_N$ (Figs. 4c,d), which tend to suggest two different YSR states as their origins.

**Similar YSR states observed on CrBr$_3$ clusters adsorbed on NbSe$_2$.** In our experiments, we occasionally find some CrBr$_3$ clusters randomly distributed on NbSe$_2$ substrate, with the typical topographic image shown in Fig. 5a. The height of these clusters is not uniform, ranging from 40 pm to 90 pm, as illustrated in Fig. 5b. This may be related to different adsorption sites and different stacking patterns of CrBr$_3$ on NbSe$_2$, which can result in different spin states of CrBr$_3$ cluster and different exchange coupling strength $J$ between the cluster and the superconductor[58,59,63,64]. A series of tunneling spectra is collected on different locations of CrBr$_3$ clusters, as displayed in Fig. 5c, presenting different kinds of in-gap states, including a pair of particle-hole symmetric conductance peaks and a nearly-zero-energy conductance peak. A hard superconducting gap recovers ~ 1 nm away from the clusters (green curve in Fig. 5c). It is considered that these various in-gap states arise from the changes of interactions between different CrBr$_3$ clusters and NbSe$_2$, and the nearly-zero-energy conductance peak is located near the QPT point of the YSR state. To confirm this, we perform $G_N$-dependent measurement for the nearly-zero-energy conductance peak, and the results are shown in Fig. 5d. With increasing $G_N$, the nearly-zero-energy conductance peak first moves to zero-energy and becomes a ZECP, which then splits into two side peaks with expanding energy separation. This further supports that the observed ZECP and the nearly-zero-energy conductance peak, no matter whether at the edge of CrBr$_3$ island or on the CrBr$_3$ clusters, are both YSR states.

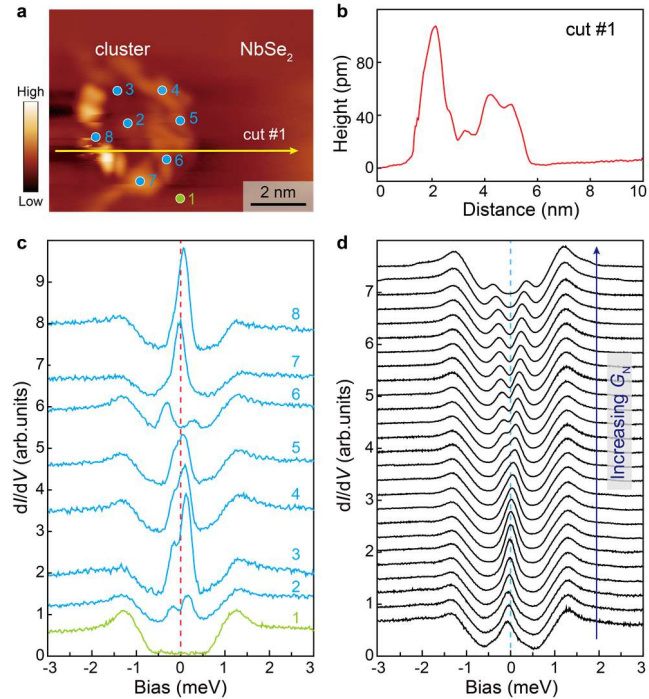

**Fig. 5 | Various YSR states observed on CrBr$_3$ clusters adsorbed on NbSe$_2$. a** Topographic image of CrBr$_3$ clusters adsorbed on NbSe$_2$ ($V_b$ = 1.0 V, $I_t$ = 30 pA). **b** Line profile taken along cut #1 in (**a**). **c** Tunneling spectra collected on different locations of CrBr$_3$ clusters. The spectrum collected on NbSe$_2$ is also listed at the bottom for comparison ($V_b$ = 3 mV, $I_t$ = 100 pA, $\Delta V$ = 0.05 mV). **d** $G_N$-dependence of the spectra with a nearly-zero-energy conductance peak ($V_b$ = 3 mV, $I_t$ = 150 - 4250 pA, $\Delta V$ = 0.05 mV).

**Discussion**

It is generally accepted that topologically protected MEMs should be spatially continuous, while the previously reported discontinuous ZECPs at the boundary of CrBr$_3$/NbSe$_2$ heterostructure are puzzling. Here by constructing the same CrBr$_3$/NbSe$_2$ heterostructure, we observe similar discrete edge states. With the advantages of a much lower sample temperature (40 mK) and higher spatial (~ 4.25 pixel/nm here compared to 1.5-2 pixel/nm in refs. 41, 47) and energy resolution (~ 50 μeV), we resolve clear lattice reconstructions of CrBr$_3$ step edges and the existence of various edge states, as well as the perfect correspondence between their spatial distribution and periods (Figs. 3g-i), which strongly suggest that the formation of edge states is closely related to the lattice reconstruction. Further tunneling transmissivity-dependent measurements demonstrate YSR states as the origin of these discrete edge states. In the junction areas of $CrBr_3$ step edge without obvious lattice reconstruction and impurities, no edge state is observed (orange curves in Fig. 3j). We suspect that lattice reconstruction may alter the arrangement patterns of $CrBr_3$ molecules, thus resulting in localized spins at the step edges. The spin state ($S_{imp}$) of the localized spins and their exchange coupling strength $J$ with $NbSe_2$ superconductor are determined by the local chemical environments. As seen from the topographic images shown in Fig. 3, Supplementary Figs. 5 and 7 of SI, the local chemical environments vary at different locations along the step edge due to the existence of defects, impurities, disorders and the arrangement patterns of $CrBr_3$ molecules; Even within a single smooth segment, they may be different for the atoms located at the boundary and inside. Therefore, different YSR states that located in different parameter space are expected at different spatial locations of the step edges, and coincidentally, they may appear at near-zero energy around $J = J_{crit}$ where they could be misinterpreted as MZMs. For different YSR states, the determination of $S_{imp}$ and $J$ is complicated and beyond the scope of this manuscript, so it will not be discussed further here.

Based on previous theory, the realization of effective topological superconducting state based on Shiba band requires strong coupling of magnetism and superconductivity[7,8,22,32,33]. And we also notice that most of the previously reported magnet-superconductor heterostructures use magnetic conductors[23-27,30,31,34-36,65], thus proximate superconductivity is induced in the magnetic layer and strongly couples with the magnetism. In our case, the $CrBr_3$ film is insulating, thus there is no superconducting proximity effect; the magnetic proximity effect is also very weak, as is revealed by the nearly same electronic states, superconducting gap structure and vortex states of bare $NbSe_2$ and $NbSe_2$ covered with $CrBr_3$ film. The weak coupling between $CrBr_3$ and $NbSe_2$ might be responsible for the absence of Shiba band and possible topological superconducting state. In the future, more strongly coupled magnet-superconductor interfaces might be explored in the search of Majorana quasiparticles along this direction.

## Methods

**Growth of $CrBr_3$ film.** $CrBr_3$ films were grown by the molecular beam epitaxy (MBE) method in the MBE chamber equipped on an ultralow-temperature STM (UNISOKU 1600). The substrates, $NbSe_2$ single crystals, were grown by chemical vapor transport (CVT) method and show a typical superconducting transition temperature of ~ 7.0 K. $CrBr_3$ single crystals were used as the evaporation sources, which were also grown by the CVT method. $NbSe_2$ single crystals were mechanically cleaved at ~ 80 K in ultrahigh vacuum with a base pressure better than $2 \times 10^{-10}$ mbar. Monolayer $CrBr_3$ films were obtained by evaporating $CrBr_3$ molecules from the Knudsen cell (350 °C) to the $NbSe_2$ substrate (270 °C), and then post-annealed at 270 °C for ~ 5 minutes to improve the film quality. Then the sample was transferred into the STM chamber for STM study.

**STM measurements.** Pt-Ir tips were used for STM measurements after being treated on a clean Au (111) substrate. The d$I$/d$V$ spectra were collected by a standard lock-in technique with a modulation frequency of 973 Hz and a modulation amplitude $\Delta V$ of 0.05-40 mV. The data in the main text were collected at the temperature of ~ 40 mK with an effective electron temperature $T_{eff}$ of ~ 170 mK (ref. 20).

## Data availability

The main data supporting the findings of this study are available within the article and its Supplementary Information files. All the raw data generated in this study are available from the corresponding author upon request.

## Code availability

All the data analysis codes related to this study are available from the corresponding author upon request.

## Acknowledgments

We thank Prof. Zhenyu Zhang, Prof. Wei Qin for helpful discussions. This work is supported by the National Natural Science Foundation of China (Grants No. 12074363 (Y.J.Y), No. 11790312 (D.L.F.), No. 12374140 (Y.J.Y), No. 11888101 (D.L.F.), No. 11774060 (Y.J.Y)), the Innovation Program for Quantum Science and Technology (Grant No. 2021ZD0302803 (D.L.F.)), the National Key R&D Program of the MOST of China (Grants No. 2023YFA1406304 (Y.J.Y)), the New Cornerstone Science Foundation (D.L.F.).

## Author contributions

CrBr$_3$ films and CrBr$_3$ single crystals were grown by Y. L., R. Y. and M. L. under the guidance of Y. Y.; NbSe$_2$ single crystals were grown by J. G.; STM measurements were performed by Y. L. and R. Y.; Data analysis was performed by Y. L., R. Y., M. L., J. G., Z. C., J. Z. and Y. Y.; Y. Y. and D. F. coordinated the whole work and wrote the manuscript. All authors have discussed the results and the interpretation.

## Competing interests

The authors declare no competing interests.